\documentstyle[12pt]{article}
\begin{document}
\baselineskip = 24pt

\title{Time Dependent Behavior of \\
Granular Material in a Vibrating Box}
\author{Jysoo Lee\\ 
Benjamin Levich Institute and Department of Physics \\
City College of the City University of New York \\
New York, NY 10031}
\date{\today}
\maketitle

\begin{abstract}

Using numerical and analytic methods, we study the time dependent
behavior of granular material in a vibrating box.  We find, by
molecular dynamics simulation, that the temporal fluctuations of the
pressure and the height expansion scale in $A f$, where $A$ ($f$) is
the amplitude (frequency) of the vibration.  On the other hand, the
fluctuations of the velocity and the granular temperature do not scale
in any simple combination of $A$ and $f$.  Using the kinetic theory of
Haff, we study the temporal behaviors of the hydrodynamic quantities
by perturbing about their time averaged values in the
quasi-incompressible limit.  The results of the kinetic theory
disagree with the numerical simulations.  The kinetic theory predicts
that the whole material oscillates roughly as a single block.
However, the numerical simulations show that the region of active
particle movement is localized and moves with time, behavior very
similar to the propagation of a sound wave.

\vspace{20pt}


\end{abstract}

\newpage
\section{Introduction}

Systems of granular particles (e.g. sand) exhibit many interesting
phenomena, such as segregation under vibration or shear, density waves
in the outflow through a hopper and a tube, and the formation of heaps
and convection cells under vibration [1-5].  These phenomena are
consequences of the unusual dynamical response of the systems, and are
for the most part still poorly understood.

We focus on the vertical vibration of a box containing granular
particles.  There are many interesting phenomena associated with this
system, such as convection cells [6-11], heap formation [12-17],
sub-harmonic instability \cite{dfl89}, surface waves \cite{pb93,mus94}
and even turbulent flows \cite{t92b}.  The basis for understanding
these diverse phenomena is the state of granular media under
vibration.  The state is characterized by the hydrodynamic fields of
the system, such as the density, velocity and granular temperature
fields.

There have been several studies on the state of granular particles
under vibration.  Thomas {\em et al.} studied the system in three
dimensions, mainly focusing on the behavior of shallow beds
\cite{tmls89}.  Cl\'{e}ment and Rajchenbach experimentally measured
the density, velocity and temperature fields of a two-dimensional
vertical packing of beads \cite{cr91}.  They found that the
temperature increases monotonically with the distance from the bottom
plate.  The same system was studied by molecular dynamics (MD)
simulation with similar results \cite{ghs92b}.  In a series of
simulations and experiments, Luding {\em et al.} studied the behavior
of the one and two-dimensional systems [25-27].  They found that the
height expansion, which is the rise of the center of mass due to the
vibration, scales in the variable $x = Af$.  Here, $A$ and $f$ are the
amplitude and the frequency of the vibration.  Warr {\it et al.}
experimentally confirmed the scaling, and they also gave an argument
for its origin \cite{whj95}.  In recent MD simulations of the
three-dimensional system, Lan and Rosato measured the density and
temperature fields \cite{lr94}.  They compared the results with the
theoretical predictions by Richman and Martin \cite{rm92}, and found
good agreement.  Also, an approximate theory was developed for the
system in one dimension, which agrees with simulations in the weak and
the strong dissipative regimes \cite{mbd94}.

In a previous paper, we studied the {\it time averaged} behavior of
the two-dimensional system using numerical and analytic methods
\cite{l95}.  Using MD simulation, we found that the time averaged
value of not only the expansion but also the density and the granular
temperature fields scale in $x$.  We also used the kinetic theory of
Haff \cite{h83} to determine the time averaged hydrodynamic fields in
the quasi-incompressible limit.  The results are, in general,
consistent with the numerical data, and in particular show scaling
behavior in the variable $x$.  We found that the origin of the scaling
can be understood within the framework of the theory.

In the present paper, we extend our study to the {\it time dependent}
behavior of the system, whose understanding is not only essential in
studying various time dependent phenomena, but also necessary to
understand the mechanisms of certain steady state phenomena.  For
example, many of the arguments for the mechanism of the convection
involve the variations of certain hydrodynamic quantities over a
vibration cycle \cite{t92a,ghs92,chi95,hyh95,l94}.

We find, by MD simulation, that the temporal fluctuations of the
pressure and the height expansion scale in $x$, while the velocity and
the granular temperature do not scale in any simple combination of $A$
and $f$.  We also study the system using the kinetic theory of Haff,
where the temporal behavior is studied by perturbing about the time
average in the quasi-incompressible limit.  The results of the kinetic
theory disagree with the numerical data.  The kinetic theory predicts
that the whole system of particles moves as one effective ``block.''
The numerical simulations, on the other hand, show that the region of
active particle movement is localized, and moves with time.  The
presence of the wave is partly responsible for the discrepancy.

The paper is organized as follows. In Sec.~\ref{sec:sim}, we specify
the interaction of the particles used in the MD simulations.  We then
present the temporal fluctuations of the expansion and the fields
obtained by the simulations.  Analytic results will be discussed in
Sec.~\ref{sec:thr}.  The continuum equations for granular material
will be given, and perturbation equations are derived.  We present the
solution of the equations, and compare with the numerical data. In
Sec.~\ref{sec:chkas}, we check the assumptions used in obtaining the
solution.  We also discuss various properties of the waves.
Conclusions are given in Sec.~\ref{sec:con}.

\section{Numerical Simulation}
\label{sec:sim}

We start by describing the interactions used in the MD simulations.
The simulations are done in two dimensions with disk shaped particles.
The interaction between the particles is that of Cundall and Strack
\cite{cs79}, which allows the particles to rotate as well as
translate.  Particles interact only by contact, and the force between
two such particles $i$ and $j$ is the following.  Let the coordinate
of the center of particle $i$ ($j$) be $\vec{R}_i$ ($\vec{R}_j$), and
$\vec{r} = \vec{R}_i - \vec{R}_j$.  We use a new coordinate system
defined by two vectors $\hat{n}$ (normal) and $\hat{s}$ (shear).
Here, $\hat{n} = \vec{r} / {\vert \vec{r} \vert}$, and $\hat{s}$ is
obtained by rotating $\hat{n}$ clockwise by $\pi/2$.  The normal
component $F_{j \to i}^{n}$ of the force acting on particle $i$ from
particle $j$ is
\begin{equation}
\label{eq:fn}
F_{j \to i}^{n} = k_n (a_i + a_j - \vert \vec{r} \vert) 
                - \gamma_n m_e (\vec{v} \cdot \hat{n}),
\end{equation}
where $a_i$ ($a_j$) is the radius of particle $i$ ($j$), and $\vec{v}
= d\vec{r}/dt$.  The first term is the linear elastic force, where
$k_n$ is the elastic constant of the material.  The constant
$\gamma_n$ of the second term is the friction coefficient of the
velocity dependent damping force, and $m_e$ is the effective mass,
$m_i m_j/(m_i + m_j)$.  The shear component $F_{j \to i}^{s}$ is given
by
\begin{equation}
F_{j \to i}^{s} = - {\rm sign} (\delta s) ~ 
{\rm min}(k_s \vert \delta s \vert, \mu \vert F_{j \to i}^n \vert).
\label{eq:fs}
\end{equation}
The term represents static friction, which requires a {\em finite}
amount of force ($\mu F_{j \to i}^{n}$) to break a contact. Here,
$\mu$ is the friction coefficient, $\delta s$ the {\em total} shear
displacement during a contact, and $k_s$ the elastic constant of a
virtual tangential spring.

The shear force also affects the rotation of the particles.  The
torque acting on particle $i$ due to particle $j$ is
\begin{equation}
\label{eq:tq}
T_{j \to i} = \vec{r}_{c} \times \hat{s} ~ F_{j \to i}^{s},
\end{equation}
where $\vec{r}_{c}$ is the vector from the center of particle $i$ to
the point where particles $i$ and $j$ overlap.  Since the particles
used in the simulations are very stiff (large $k_{n}$), the area of
the overlap is very small.  It is thus a good approximation to use
$-a_{i} \hat{n}$ as $\vec{r}_{c}$.

A particle can also interact with a wall.  The force and torque on
particle $i$, in contact with a wall, are given by (\ref{eq:fn}) -
(\ref{eq:tq}) with $a_j = 0$ and $m_e = m_i$.  A wall is assumed to be
rigid, i.e., it is not moved by collisions with particles. Also, the
system is under a gravitational field $\vec{g}$. A more detailed
explanation of the interaction is given elsewhere \cite{lh93}.

The movements of the particles are calculated using a fifth order
predictor-corrector method.  We use two Verlet tables.  One is a usual
table with finite skin thickness.  The other table is a list of pairs
of {\it actually} interacting particles, which is needed to calculate
the shear force.  The interaction parameters used in this study are
fixed as follows, unless otherwise specified: $k_n = 10^5, k_s = 10^5,
\gamma _n = 2 \times 10^2$ and $\mu = 0.2$. The timestep is taken to
be $5 \times 10^{-6}$. This small timestep is necessary for the large
elastic constant used in the simulations.  For too small values of the
elastic constant, the system loses the character of a system of
distinct particles, and behaves like a viscous material.  In order to
avoid artifacts of a monodisperse system (e.g., hexagonal packing), we
choose the radius of the particles from a Gaussian distribution with
the mean $0.1$ and the width $0.02$.  The density of the particles is
$0.1$.  Throughout this paper, CGS units are implied.

We put the particles in a two-dimensional rectangular box.  The box
consists of two horizontal (top and bottom) plates which oscillate
sinusoidally along the vertical direction with given amplitude $A$ and
frequency $f$.  The separation between the two plates $H$ is chosen to
be much larger ($10^5$ times) than the average radius of the
particles, so the particles do not interact with the top plate for all
cases studied here.  We apply a periodic boundary condition in the
horizontal direction.  The width of the box is $W = 1$.  We also try
different values of $W$, and find no essential difference in the
following results.

We start the simulation by inserting the particles at random positions
in the box. We let them fall by gravity and wait while they lose
energy by collisions. We wait for $10^5$ iterations for the particles
to relax, and during this period we keep the plates fixed.  The
typical velocity at the end of the relaxation is of order $10^{-2}$.
After the relaxation, we vibrate the plates for $50$ cycles before
taking measurements in order to eliminate any transient effect.
Measurements are made during the next $200$ cycles.

We measure hydrodynamic quantities---density, velocity and granular
temperature---which characterize the state of the system.  The most
detailed information is contained in the time series of their fields
(e.g., density field), which will be discussed later.  Here, we want
to start with something simple and representative of the system.

The center of mass of the particles can be loosely related to the
density.  Let $y(t)$ be the vertical coordinate of the center of mass
at time $t$.  The mean density is related to the mean interparticle
distance, which is also related to $y(t)$.  Since $y(t)$ is a scalar
which also can be easily measured, we study it as a representative of
the density.  In the same spirit, we study the space averaged vertical
velocity $V_y(t)$ and granular temperature $\tau(t)$ instead of the
complete local fields.  We define the spatial average of $A(x,y,t)$ as
\begin{equation}
\langle A(t) \rangle = {\int \int A(x,y,t) \rho(x,y,t) dx dy \over
                        \int \int \rho(x,y,t) dx dy},
\end{equation}
where $\rho(x,y,t)$ is the density field at position $(x,y)$.  It thus
follows that $V_y(t) = \langle v_y(t) \rangle$ and $\tau(t) = \langle
T(t) \rangle$, where $v_y(x,y,t)$ and $T(x,y,t)$ is the vertical
velocity and the granular temperature field, respectively.  Another
quantity of interest is the pressure field $p(x,y,t)$.  For
information on this quantity, we will study the total pressure at the
bottom
\begin{equation}
p_o(t) = {\int p(x,y=0,t) \rho(x,y=0,t) dx \over 
               \int \rho(x,y=0,t) dx}.
\end{equation}

We studied the time averaged behavior of the system in our previous
paper \cite{l95}.  Here, we study the temporal behavior, especially
focusing on the variations of the fields within a vibration cycle.  We
first measure the temporal fluctuation, which is defined as the
standard deviation of a temporal sequence.  Let $\langle A \rangle
_{t}$ be the time average of $A(t)$; then, we use
\begin{eqnarray}
\bar{y}_{\rm exp} = \langle y_{\rm exp} \rangle _t, &
\delta y_{\rm exp} = \sqrt{\langle y_{\rm exp}^2 \rangle_t - 
\langle y_{\rm exp} \rangle_t^2} \nonumber \\
\bar{V}_y = \langle V_y \rangle _t, &
\delta V_y = \sqrt{\langle V_y^2 \rangle_t - \langle V_y \rangle_t^2}
\nonumber \\
\bar{\tau} = \langle \tau \rangle _t, &
\delta \tau = \sqrt{\langle \tau^2 \rangle_t - \langle \tau \rangle_t^2}
\nonumber \\
\bar{p}_o = \langle p_o \rangle _t, &
\delta p_o = \sqrt{\langle p_o^2 \rangle_t - \langle p_o \rangle_t^2},
\end{eqnarray}
where the expansion $y_{\rm exp}(t)$ is defined as the difference
between $y(t)$ during and before the vibration.  We measure these
quantities at every $1/100$ of a period for $200$ cycles.

In our previous paper, it was shown that $\bar{y}_{\rm exp}$ scales in
$A f$, in agreement with earlier simulations and experiments [25-28].
We find that the rotation of the particles included in the present
simulation does not change this scaling, but does significantly
decrease the value of $\bar{y}_{\rm exp}$ \cite{lu95}.  The decrease is
probably due to the fact that the average translational energy becomes
smaller, since some of the energy is transferred to the rotation.  We
then consider the fluctuation of the expansion $\delta y_{\rm exp}$.
In Fig.~1(a), we show $\delta y_{\rm exp}$ for several values of $A$
and $f$.  The quality of the scaling is not very good, but the data is
still consistent with $A f$ scaling, especially without the persistent
deviation at low $A$ part of the $f = 20$ data.  This deviation is
also present in the scaling of $\bar{y}_{\rm exp}$.

The behavior of $\tau(t)$ is very similar.  It was shown that
$\bar{\tau}$ scales in $A f$ \cite{l95}.  Again, we find that the
rotation does not change the scaling of $\bar{\tau}$, but does change
its value.  The situation becomes a little different when we consider
the fluctuation $\delta \tau$ as shown in Fig.~1(b).  It is clear from
the figure that $\delta \tau$ does not scale in $A f$.  In fact, it
does not scale in any of the simple combinations of $A$ and $f$ we
have tried.

In the previous paper, the behaviors of $V_y(t)$ and $p_o(t)$ were not
discussed in detail, since their time averaged values are trivial.
Since the system is in a steady state, $\bar{V}_y$ is zero, which is
also confirmed by the simulations.  Since the pressure $p_o(t)$ is
caused by the weight of the particles, one might guess $\bar{p}_o$ is
simply the total weight of the particles divided by the area of the
bottom ($W = 1$).  We find that $\bar{p}_o$ is indeed a constant
independent of $A$ and $f$, whose value is consistent with the total
weight.  On the other hand, the behavior of their fluctuations is far
from trivial.  The fluctuation of the velocity $\delta V_y$ for
several values of $A$ and $f$ is shown in Fig.~1(c).  It is apparent
that $\delta V_{y}$ does not scale in $A f$.  Also, $\delta V_{y}$
does not scale in any simple combination of $A$ and $f$.  The data for
the fluctuation of the pressure $\delta p_o$ is consistent with
scaling in $A f$ as shown in Fig.~1(d).  The quality of collapse is
again not very good, especially for the low $A$ part of the $f = 20$
data.

The behavior of the time averaged quantities is either trivial
($\bar{V}_{y}$ and $\bar{p}_o$), or scales in $A f$ ($\bar{y}_{\rm
exp}$ and $\bar{\tau}$).  The reason for the trivial behavior has been
discussed, and the scaling in $A f$ can be understood from a kinetic
theory of granular particles \cite{l95}.  The behavior of the
fluctuations is, however, not easy to understand.  For example, one
might naively expect that $\delta V_{y}$ behaves as $A f$, the
velocity fluctuation of the bottom plate; but the numerical data
suggests this is not so.  Also, one might guess $\delta p_o$ behaves
as $A f^{2}$, the variation of the effective gravity; but the
numerical simulations suggest scaling in $A f$.  Thus, the behaviors
of the temporal fluctuations seem to be inconsistent with an intuitive
picture.  In the next section, we discuss an attempt to understand
this behavior.

\section{Kinetic Theory}
\label{sec:thr}

In this section, we study the time dependent behavior of the system
using a kinetic theory of granular material.  We use the formalism by
Haff \cite{h83}, which was successfully applied to the time averaged
behavior of the system \cite{l95}.  For more details on the formalism
and other kinetic theories of granular material, see Ref.~\cite{l95}
and references therein.

Haff's formulation consists of equations of motion for mass, momentum
and energy conservation. The mass conservation equation is
\begin{equation}
\label{eq:mass}
{\partial \over \partial t} \rho + \vec{\nabla} \cdot (\rho \vec{v}) 
= 0,
\end{equation}
where $\rho$ and $\vec{v}$ are the density and the velocity fields,
respectively.  Next is the $i$-th component of the momentum
conservation equation,
\begin{equation}
\label{eq:mom}
\rho {\partial \over \partial t} v_i + \rho (\vec{v} \cdot 
\vec{\nabla}) v_i = {\partial \over \partial x_i} [ -p + \lambda 
(\vec{\nabla} \cdot \vec{v})] + {\partial \over \partial x_j} [\eta
({\partial v_j \over \partial x_i} + {\partial v_i \over \partial 
x_j})] + \rho g_i,
\end{equation}
where summation over index $j$ is implied. The coefficients $\lambda$
and $\eta$ are viscosities which will be determined later. Also, $p$
is the internal pressure, and $g_i$ is the $i$-th component of the
gravitational field. Although (\ref{eq:mom}) resembles the
Navier-Stokes equation, the coefficients as well as the internal
pressure are now functions of the fields instead of being
constant. The last of the equations of motion is energy conservation,
\begin{eqnarray}
\label{eq:energy}
{\partial \over \partial t} ({1 \over 2} \rho v^2 + {1 \over 2} 
\rho T) & + & {\partial \over \partial x_i} 
[( {1 \over 2} \rho v^2 + {1 \over 2} \rho T) v_i] \\ \nonumber 
& = & -{\partial \over \partial x_i} (p v_i) \\ \nonumber
& + & {\partial \over \partial x_i}[\lambda (\vec{\nabla} \cdot 
\vec{v}) v_i] + {\partial \over \partial x_i}[\eta ({\partial v_i 
\over \partial x_j} + {\partial v_j \over \partial x_i}) v_j] \\ 
\nonumber & + & \rho v_i g_i \\ \nonumber 
& + & {\partial \over \partial x_i} [K {\partial \over 
\partial x_i} ({1 \over 2} \rho T)] - I.
\end{eqnarray}
Here, $T$ is the granular temperature field, $K$ is the ``thermal
conductivity,'' $I$ is the rate of the dissipation due to inelastic
collisions, and summations over indices $i$ and $j$ are implied.
Although the form of (\ref{eq:energy}) is somewhat different from that
of the Navier-Stokes equations, the equation can still be easily
understood. The left hand side of (\ref{eq:energy}) is simply the
material derivative of the total kinetic energy, where the total
kinetic energy is divided into the convective part (involving
$\vec{v}$) and the fluctuating part (involving $T$). On the right hand
side of the equation, the first three lines are simply the rate of
work done by the internal pressure, viscosity and gravity,
respectively. The term involving $K$ is the rate of energy transported
by ``thermal conduction.'' The dissipation term $I$, which is a
consequence of the inelasticity of the particles, is responsible for
many of the unique properties of granular material.

We now discuss the coefficients which are yet to be determined.
Derivation of the relations of these coefficients to the fields is
based on intuitive arguments \cite{h83}. Also, the derivation assumes
that the density is not significantly smaller than the close-packed
density, i.e., the system is almost incompressible. The relation for
the internal pressure is
\begin{equation}
\label{eq:intp} 
p = t d \rho {T \over s},
\end{equation}
where $t$ is an undetermined constant, and $d$ is the average diameter
of the particles. The variable $s$, which is roughly the gap between
the particles, is related to the density by
\begin{equation}
\label{eq:sandr}
\rho \equiv {m \over (d + s)^3},
\end{equation}
where $m$ is the average mass of the particles. Then, the viscosity
$\eta$ is given as
\begin{equation}
\label{eq:eta}
\eta = q d^2 \rho {\sqrt{T} \over s},
\end{equation}
where $q$ is an undetermined constant. In a similar way, the thermal
conductivity is found to be
\begin{equation}
\label{eq:K}
K = r d^2 {\sqrt{T} \over s}.
\end{equation}
Here again, $r$ is an undetermined constant. Finally, the rate of
dissipation is
\begin{equation}
\label{eq:I}
I = \gamma \rho {T^{3/2} \over s},
\end{equation}
where $\gamma$ is an undetermined constant.  The viscosity $\lambda$
is left undetermined, due to the fact that, in the range where these
relations are valid, the term containing $\lambda$ is negligible and
is dropped from the calculation.

We impose two constraints in order to make the equations analytically
tractable.  The first is the horizontal periodic boundary condition.
Due to the boundary condition, there are no significant variations of
the fields along the horizontal direction.  Thus, we only have to deal
with a one-dimensional equation, instead of a two or three-dimensional
one.  The other constraint is incompressibility, which is a little
tricky.  Incompressibility implies, strictly speaking, that the
density $\rho$ is constant. Due to the relation between $\rho$ and $s$
(\ref{eq:sandr}), $s$ also has to be constant.  Here, we are
interested in the situation where $s$ is much smaller than $d$, but
still non-zero. In such a case, the variation of the density can be
ignored, but not the variation of a variable that depends directly on
$s$. We call this condition quasi-incompressibility.

Under these conditions, we solved the equations for the time averaged
fields,
\begin{eqnarray}
T^{(0)}(y) & \simeq & B^{2} ~ v_w^{2} ~ {y \over y_o - y} ~ \exp 
(-2y/\ell) \nonumber \\
s^{(0)}(y) & \simeq & {t^2 d^2 \rho_o \over g}~ B^2 ~ v_w^2 ~{y_o 
\over (y_o - y)^2} ~ \exp (-2y / \ell) \nonumber \\
v_{y}^{(0)}(y) & = & 0 \nonumber \\
p^{(0)}(y) & = & \rho_o g (y_o - y).
\end{eqnarray}
Here, $v_{w} = 2 \pi A f$ the maximum velocity of the bottom, $y_{o}$
the height of the free surface, and $\ell = \sqrt{r / \gamma}~d$ the
dissipation length.  Also, $\rho _{o}$ is the density of the maximum
packing, and 
\begin{equation}
B^{2} = {(1 + e_w)^2 \over 1 - e_w^2 - (2rd / a\ell) ~ (1 - \ell / 
2y_o)},
\end{equation}
where $e_w$ is the coefficient of restitution of collisions between a
particle and a wall \cite{l95}.

We study the time dependent behavior of a quantity by perturbing it
from its time averaged value.  We assume that the perturbation term
oscillates with $f$---the frequency of the vibration.  In general, the
assumption is not valid, since one has to consider all the modes with
different frequencies.  However, when the amplitude of the vibration
is small enough, in many cases, the mode with frequency $f$ dominates
the time dependent behavior.  We expect there is a range of $A$ in
which the assumption is valid, which will be determined later by the
numerical simulations.  We thus use
\begin{eqnarray}
\label{eq:pertvs}
T(y,t) & = & T^{(0)}(y) + T^{(1)}(y) \cdot \exp (i \omega t) \nonumber \\
s(y,t) & = & s^{(0)}(y) + s^{(1)}(y) \cdot \exp (i \omega t) \nonumber \\
v_{y}(y,t) & = & v_{y}^{(0)}(y) + v_y^{(1)}(y) \cdot \exp (i \omega t) 
\nonumber \\
p(y,t) & = & p^{(0)}(y) + p^{(1)}(y) \cdot \exp (i \omega t).
\end{eqnarray}

Substituting (\ref{eq:pertvs}) into the mass conservation condition
(\ref{eq:mass}) and using the quasi-incompressibility condition, we
obtain $d v_y^{(1)}(y) / d y = 0$.  Here and in the rest of the
calculation, we consider terms up to the first order in the
expansion. Since $v_{y}(y=0,t)$ should be the velocity of the bottom
plate,
\begin{equation}
\label{eq:pvel}
v_y^{(1)}(y) = i v_w,
\end{equation}
which is a consequence of quasi-incompressibility and the
one-dimensional nature of the system.  Also, momentum conservation
equation (\ref{eq:mom}), combined with (\ref{eq:pertvs}), becomes
\begin{equation}
\label{eq:ppress}
p^{(1)}(y) = \rho_o A \omega ^2 (y_o - y),
\end{equation}
which can be easily understood.  The pressure at the bottom is
proportional to the total weight of the particles.  Thus, the
fluctuation of $p_o(t)$ is the total mass $\rho _o y_o$ times the
fluctuation of the effective gravity $A \omega^2$.  Finally, we
consider the energy conservation equation (\ref{eq:energy}).  Combined
with (\ref{eq:pertvs}), it becomes
\begin{eqnarray}
\label{eq:pfluc}
i \omega \rho_o U^{(0)} U^{(1)} & + & i \rho_o v_y^{(1)}
U^{(0)} {d \over dy} U^{(0)} \nonumber \\
& = & {r d \over t} {d \over dy} (p^{(0)} {d \over dy} U^{(1)} +
p^{(1)} {d \over dy} U^{(0)})  \nonumber \\
& - & {\gamma \over td} (U^{(1)} p^{(0)} + U^{(0)} p^{(1)}),
\end{eqnarray}
where we introduce the variable $U(y,t) = \sqrt{T(y,t)}$.  One has to
solve (\ref{eq:pfluc}) for $T(y,t)$.  We can not, unfortunately, get
an analytic solution of the resulting nonlinear differential equation.
Since $s(y,t)$ has to be calculated from the relation (\ref{eq:intp})
between $T(y,t), s(y,t)$ and $p(y,t)$, we also can not get an
expression for $s^{(1)}(y)$.

We compare the results from the kinetic theory with the numerical
simulations (Fig.~1).  First, we consider $v_{y}(y,t)$. Since the
temporal fluctuation of $v_{y}(y,t)$ is proportional to
$v_{y}^{(1)}(y)$, the kinetic theory predicts $\delta V_{y} \sim A f$.
The data from the simulations, however, is inconsistent with this
scaling.  The situation is similar for $p(y,t)$.  The kinetic theory
predicts $\delta p_o \sim A f^{2}$, while the numerical data scales in
$A f$.  We do not have analytic expressions for $T(y,t)$ and $s(y,t)$
to compare with the simulation data.

The failure of the kinetic theory, when applied to the time dependent
behavior, is in sharp contrast with its success in studying the time
averaged behavior.  The kinetic theory correctly predicts the scaling
behavior of all the time averaged hydrodynamic quantities.  We suspect
the failure is associated with the breakdown of a key assumption(s)
used in the theory.  Three key assumptions, besides the ones employed
in the formulation of the kinetic theory, are made to obtain a simple
analytic solution for the time dependent behavior.  The first is
quasi-incompressibility, which is shown to be valid for $A f \ll 1.5$
from the study of the time averaged behavior \cite{l95}.  However, the
scaling behavior of the time averaged quantities remains unchanged
even in the compressible regime.  The second assumption is that the
time dependence of a quantity is a sinusoidal oscillation with
frequency $f$, which we expect to be valid for small $A$.  The last
assumption is that the time dependent term in (\ref{eq:pertvs}) is
much smaller than its time averaged value in order for the
perturbation to be valid.  In the next section, we check the validity
of these assumptions by comparing to the numerical simulations.

\section{Validity of Assumptions}
\label{sec:chkas}

First, we check the assumption that the mode with the driving
frequency dominates the time dependent behavior of the hydrodynamic
quantities.  We obtain a time series of $y_{\rm exp}(t)$ by measuring
it at every $1/100$ of a period for $200$ cycles.  We then calculate
the power spectrum of $y_{\rm exp}(t) - \bar{y}_{\rm exp}$ using a FFT
routine in the NAG library (c06gbf).  The results with $f = 100$ are
shown in Fig.~2.  The mode with $f = 100$ is dominant for $A = 0.01$
(Fig.~2(a)).  When $A$ is increased further to $0.03$, however, the $f
\simeq 20$ mode becomes dominant (Fig.~2(b)).  The results with $f =
20$ are entirely similar.  The $f = 20$ mode loses its dominance when
$A$ is increased to about $0.5$.  In both cases, the mode with
frequency $f$ is dominant until $\Gamma \sim 10$, where $\Gamma = A
\omega ^2 / g$.

The measurements of $V_y(t)$ and $p_{o}(t)$ also support the above
observations.  In Fig.~3(a), we show the variation of $V_y(t)$ in one
cycle, where $f = 100$ and the data are averaged over $200$ periods.
The curve with $A = 0.01$ is nearly sinusoidal, while clear deviation
is seen for $A = 0.03$.  An important point to note is the behavior of
the maximum value of $V_y(t)$.  The kinetic theory predicts, in
(\ref{eq:pvel}), the maximum value to be proportional to $A$, which is
clearly not consistent with the data.  The measured value is quite
smaller than what is predicted.  For example, for $A = 0.03$, the
predicted value is $6 \pi$, while the measured value is about $2.4$.
In Fig.~3(b), we show $p_{o}(t)$ in one period, where again $f = 100$
and the data is averaged over $200$ cycles.  The curve seems to
deviate from the sinusoidal even for small $A$, and it is difficult to
determine whether the mode with $f = 100$ dominates.  Again, the
predicted behavior of the maximum value of $p_{o}(t)$ is not
consistent with the measurement.  The maximum value of $p_{o}(t)$ is
predicted to increase linearly with $A$, as in (\ref{eq:ppress}),
which is clearly not consistent with the data.  The observed maximum
value ($\sim 400$) is quite smaller than the predicted value ($\sim
4,000$).

The mode with the driving frequency dominates the time dependent
behavior for small values of $A$, where a rough criterion for the
dominance is $\Gamma \ll 10$.  Also, it was shown that
quasi-incompressibility is valid for $A f < 1.5$ \cite{l95}.  The
validity of the linear perturbation approximation (\ref{eq:pertvs}) is
a little tricky.  We require that the perturbation terms are smaller
than the time averaged terms, which is valid for the expansion and the
temperature.  The two terms are, however, comparable for the pressure
even at small value of $A = 0.05$.  The consequence of the large
pressure fluctuation on the validity of the perturbation is unclear.
For large $A$, all of the above assumptions are not valid, which
complicates the analysis of the system.  For example, in order to
study the time dependent behavior, one has to consider additional
modes with different frequencies.

The surprise is that the predictions for the maximum values of the
vertical velocity and the pressure are not correct even when all the
assumptions seem to be valid (e.g., $A = 0.01$ and $f = 100$).  We
inspect again the predictions of the kinetic theory.  As given in
(\ref{eq:pvel}), the velocity field is uniform, and its value is $v_w$
the velocity of the bottom.  Therefore, the solution of the
perturbation expansion suggests that the whole system of particles is
moving as a single ``block'' attached to the bottom.  The spatial and
temporal variations of the other fields do not change the single block
picture, but rather describe the structure of the block.  A
consequence of the picture is that the pressure at the bottom is
proportional to the effective gravity which reaches its maximum
$(\Gamma + 1)g$ at phase $3 \pi / 2$ of the vibration, which is
exactly (\ref{eq:ppress}).

However, the measurements of $V_{y}(t)$ and $p_o(t)$ are not
consistent with the picture.  The measured maximum value of $V_y(t)$
is much smaller than what is predicted, which suggests that only a
small fraction of the particles move together at a given time.  Also,
the fact that the measured maximum value of $p_o(t)$ is smaller than
the prediction also supports this observation.  Furthermore, the phase
at which $p_o(t)$ reaches the maximum is about $0$, in contrast to $3
\pi / 2$ suggested by the single block picture.  The discrepancy can
be understood as follows.  Since the maximum acceleration of the
bottom is larger than that of gravity, particles initially lying on
the bottom will be ``launched'' at a certain phase of the vibration.
The maximum pressure at the bottom will occur when most of the
launched particles come back and collide with the bottom, which occurs
around $\phi = 0$.

The direct evidence against the single block picture is the time
evolution of the whole velocity field shown in Fig.~4(a).  Here, we
use $f = 100, A = 0.01$, at which the assumptions used to derive the
predictions of the kinetic theory seem to be valid.  It is clear from
the figure that the region of significant motion is localized, and
travels like a wave.  The propagation of the disturbance seems to be
very similar to that of sound waves in a gas.  Also, the maximum
velocity is about $6$, close to the prediction $2 \pi$.  The
localization of the particle motion can also be seen in the time
evolution of the granular temperature field shown in Fig.~4(b).
Again, the region of high temperature is localized, and travels
upwards \cite{gh92,aa95}.  Furthermore, the location of the high
temperature region coincides with that of the large velocity region.
The density field, on the other hand, does not vary significantly as
shown in Fig.~4(c), which agrees with the previous experiment
\cite{cr91}.  It is clear that the presence of the ``waves'' changes
the behaviors of the fields, and possibly their scaling properties.
The absence of the waves in the single block solution is, at least,
partly responsible for the failure of the kinetic theory.  The absence
is due to quasi-incompressibility.  In fact, it is easy to derive the
single block picture only from the quasi-incompressibility condition
and one-dimensional nature of the equation.  It is thus necessary to
consider the general case of the kinetic theory of a compressible gas,
which unfortunately is quite complicated.

We want to finish this section by discussing some properties of the
waves.  We first consider the motion of the maximum disturbance.  In
Fig.~5(a), we show the phase $\phi_{\rm max} (y)$ at which $v_{\rm
y}(y,t)$ reaches a maximum with $f = 100$ and several values of $A$.
In other words, we plot the position of the maximum velocity in the $y
- \phi$ plane.  The velocity of the wave, inversely proportional to
the slope of $\phi _{\rm max}(y)$ curve, is a bit small.  The time
needed for the wave to propagate from the bottom to the top of the
pile is of the order of the period of the vibration.  The simulations
with $f = 50$ show that, for the same values of $A$, the velocity of
the wave does not change significantly, suggesting that there is a
fixed time scale for the wave propagation.  Also, it can be seen from
the figure that the velocity decreases when either $y$ or $A$
increases with the other parameters fixed.  The decrease probably
results from the decrease of the collision frequency between
particles, due to the decrease of the density.  Also, the location of
the maximum granular temperature in the $y - \phi$ plane is shown in
Fig.~5(b), where one can see the close correlation with Fig.~5(a).  In
fact, the maximum temperature always occurs just above the maximum
velocity at a given phase, which probably is the point of the largest
velocity gradient.  We now discuss the values of the maximum
disturbances.  In Fig.~6(a), we show the maximum value of $v_{\rm
y}(y,t)$ for given $y$---$v_{\rm y}(y,\phi _{\rm max}(y))$.  The
maximum value decreases with $y$ roughly as an exponential, and the
rate of the decrease is larger for larger $A$ with fixed $f$.  The
decrease is due to two causes: (1) some of the vertical velocity
component is transferred to the horizontal one by interparticle
collisions; (2) the kinetic energy is lost by inelastic collisions.
The results with $f = 20$ are essentially the same.  Also, the
behavior of the value of the maximum temperature $T(y,\phi _{\rm
max}(y))$, as shown in Fig.~6(b), is very similar to that of $v_{\rm
y}(y,\phi _{\rm max}(t))$.  It decreases roughly as an exponential,
and it decays faster with larger $A$ with fixed $f$, where the origin
of the decrease is the same as the velocity field.

\section{Conclusion}
\label{sec:con}

We have studied the temporal behavior of granular material in a
vibrating box.  We find that the temporal fluctuations of $y_{\rm
exp}(t)$ and $p_{o}(t)$ scale in $A f$, while no scaling is found for
the fluctuations of $\tau (t)$ and $V_y(t)$.  We study the behavior
using the kinetic theory of Haff, where we perturb the hydrodynamic
quantities from their time averaged values.  The results of the
kinetic theory are not consistent with the numerical data.  We argue
that the failure of the theory is, at least, partly due to the waves
found in the simulations.

We discuss some possible ways to study the time dependent behavior of
the system.  Since the main problem of the present theory, as
discussed above, is the condition of quasi-incompressibility, it
sounds reasonable to study the system of equations in a fully
compressible regime.  However, the system of equations becomes too
complicated.  The three conservation equations
(\ref{eq:mass})-(\ref{eq:energy}) are written in general form, and do
not need any modification. The relations of $p, \eta, K, I$ to the
fields (\ref{eq:intp})-(\ref{eq:I}) as well as the boundary conditions
have to be modified. It is not the modifications themselves, but the
complexity of the resulting equations, that makes an analytic solution
too difficult to obtain.  However, we can still gain some information
about the system by a perturbative or an approximate method, as well
as the numerical solution of the kinetic equations.

We thank Joel Koplik for many useful discussions and comments on the
manuscript, and Michael Tanksley for critical reading of the
manuscript. This work is supported in part by the Department of Energy
under grant DE-FG02-93-ER14327.

\newpage
\section*{Figure Captions}

\begin{description}

\item [Fig.~1:] Scaling behaviors of the hydrodynamic quantities.  Each
datum is averaged over at least $3$ samples, where $20 000$
measurements are made in a sample. (a) The fluctuation of the
expansion $\delta y_{\rm exp}$ seems to scale in $A f$, where (b) the
fluctuation of the temperature $\delta \tau$ does not scale.  (c) The
fluctuation of the velocity $\delta V_y$ does not scale, where (d) the
fluctuation of the pressure $\delta p_o$ seems to scale in $A f$.

\item [Fig.~2:] Power spectrum of $y_{\rm exp}(t)$ with $f = 100$ and
(a) $A = 0.01$, (b) $A = 0.03$.  The expansion is measured at every
$10^{-4}$ second for $200$ cycles.  The mode with $f = 100$ is
dominant at $A = 0.01$, but loses its dominance at $A = 0.03$.

\item [Fig.~3:] Time evolution of (a) $V_{y}(t)$, (b) $p_o(t)$ in one 
cycle, where $f = 100$ and the data are averaged over $200$ cycles.
Deviation from a sinusoidal is apparent for $A = 0.03$.

\item [Fig.~4:] Time evolution of (a) $v_{y}(y,t)$, (b) $T(y,t)$ and
(c) $\rho (y,t)$ fields, where $f = 100, A = 0.01$ and the data are
averaged over $200$ cycles.  The density field is normalized to be the
volume fraction.

\item [Fig.~5:] The position of (a) the maximum velocity and (b) the
maximum temperature in the $y - \phi$ plane with $f = 100$ and several
values of $A$.  The data are averaged over $200$ cycles.  Note that
the two sets of the curves are almost identical.

\item [Fig.~6:] The maximum value of (a) the vertical velocity $v_{\rm
y}(y,\phi _{\rm max}(y))$ and (b) the temperature $T(y,\phi _{\rm
max}(y))$ at height $y$ with $f = 200$.  The data are averaged over
$200$ cycles.

\end{description}

\end{document}